\documentclass[nameyear]{elsart}

\begin{document}

\newcommand{\lesssim}{\mathrel{
\hbox{\rlap{\hbox{\lower4pt\hbox{$\sim$}}}\hbox{$<$}}}}
\newcommand{\gtrsim}{\mathrel{
\hbox{\rlap{\hbox{\lower4pt\hbox{$\sim$}}}\hbox{$>$}}}}

\newcommand{\bx}{{\vec x}}
\newcommand{\bs}{{\vec s}}
\newcommand{\br}{{\vec r}}
\newcommand{\bq}{{\vec q}}
\newcommand{\bb}{{\vec b}}

\newcommand{\GHz}{~{\rm GHz}}
\newcommand{\mJy}{~{\rm mJy}}
\newcommand{\micas}{~\mu\,{\rm a.s.}}
\newcommand{\mas}{~{\rm m.a.s.}}
\newcommand{\s}{~{\rm s}}
\newcommand{\hr}{~{\rm hr}}
\newcommand{\dy}{~{\rm d}}
\newcommand{\cm}{~{\rm cm}}
\newcommand{\m}{~{\rm m}}
\newcommand{\kpc}{~{\rm kpc}}
\newcommand{\Gpc}{~{\rm Gpc}}
\newcommand{\kms}{~{\rm km~s}^{-1}}

\newcommand{\apj}{ApJ}
\newcommand{\apjs}{ApJS}
\newcommand{\araa}{ARA\&A}
\newcommand{\aap}{Astr. \& Ap.}
\newcommand{\mnras}{MNRAS}

\newcommand{\eg}{{\it e.g.~}}
\newcommand{\ie}{{\it i.e.~}}

\begin{frontmatter}

\title{Radio scintillation of gamma-ray-burst afterglows
\thanksref{journal}}
\thanks[journal]{submitted to \emph{New Astronomy}}

\author{Jeremy Goodman\thanksref{email}}

\address{Princeton University Observatory, Peyton Hall,
    Princeton, NJ 08544}

\thanks[email]{E-mail: jeremy@astro.princeton.edu}

\begin{abstract}

Stars twinkle to the eye through atmospheric turbulence, but planets,
because of their larger angular size, do not.  Similarly,
scintillation due to the local interstellar medium will modulate the
radio flux of gamma-ray-burst afterglows and may permit indirect
measurements of their angular sizes.  The amplitude of
refractive scintillation is of order ten percent at ten
gigahertz unless the source size is much larger than the expected
size, of order ten microarcseconds.  Diffractive scintillation is
marginally possible, depending sensitively on the source size,
observing frequency, and scattering measure of the interstellar
medium.\\

\end{abstract}
\end{frontmatter}

\clearpage

\section{INTRODUCTION}

At the time of writing, three gamma-ray bursts discovered by the
BeppoSAX satellite have been associated with transient sources
at longer wavelengths (\cite{IAU6572}, \cite{IAU6610}, \cite{IAU6617},
\cite{IAU6649}, \cite{IAU6654}).
Absorption lines of iron and magnesium at $z=0.835$ in the optical
counterpart of GRB0508 appear to prove that this burst, at least, 
lies at a cosmological distance (\cite{IAU6655}).
GRB0508 is also the first burst associated with a
radio transient (\cite{IAU6662}, \cite{IAU6670}).

These Xray-to-radio ``afterglows''
are broadly consistent with models based on
relativistic blastwaves at cosmological distances
(\cite{RM92}, \cite{PR93}, \cite{W97}, \cite{WRM},
and references therein).
In these models, the shock Lorentz factor is $\gtrsim 300$
at the time of the gamma-ray event
and falls thereafter as $\Gamma_{\rm s}\propto t^{-3/8}$,
$t$ being the time measured at Earth.
The bulk Lorentz factor of the postshock material is
$\Gamma_{\rm ps}= 2^{-1/2}\Gamma_{\rm s}$.
For synchrotron emission,
the flux density $F_\nu$ at frequency $\nu$ is predicted to peak
at a time $t_\nu\propto \nu^{-3/2}$ if the source is
optically thick (\cite{PR93}), or
$t_\nu\propto \nu^{-2/3}$ if optically thin
(cf. \cite{W97}, \cite{WRM}).

A lower limit to the angular size $\theta_s=r_s/D$ of the radio
source follows if the brightness temperature
is not larger than the Compton limit for incoherent synchroton radiation,
$T_b\lesssim\Gamma_{\rm ps}(1+\beta_{\rm ps})T_{\max}$ with
$T_{\max}\approx 10^{12}~\mbox{K}$ (\cite{KPT}):
\begin{equation}
\theta_s\ge 1.5~\left[\frac{1+z}{(1+\beta_{\rm ps})\Gamma_{\rm ps}}
\right]^{1/2}
\left(\frac{F_\nu}{\mJy}\right)^{1/2}
\nu_{\rm 10}^{-1/2}\left(\frac{T_{\max}}{10^{12}K}\right)^{-1/2}\micas,
\label{eq:sync}
\end{equation}
where $\nu_{10}\equiv\nu/(10\GHz)$.
\cite{Sari} has pointed out that 
the radius of the shock at observed time $t$
is $R_{\rm s}=8\Gamma^2_{\rm s}ct$ for $\Gamma_{\rm s}\gg 1$
since $\Gamma_{\rm s}\propto t^{-3/8}$.
Most of the emission is seen from a disk of radius 
$R_{\rm s}/\Gamma_{\rm ps}$, so
\begin{equation}
\theta_{\rm s} = \frac{ct}{d}\times\cases{
	8\sqrt{2} \Gamma_{\rm s} & $\Gamma_{\rm s}\gg 1$;\cr
	(5/2)\beta_{\rm s} & $\beta_{\rm s}\equiv
		(1-\Gamma_{\rm s}^{-2})^{1/2}\ll 1$\cr}
\label{eq:super}
\end{equation}
where $d$, the proper motion distance of the source, is $(1+z)$ times 
the angular-size distance (cf. \cite{Weinberg}),
and the nonrelativistic formula above follows from the
Sedov solution.
Comparison of relations (\ref{eq:sync}) and (\ref{eq:super})
yields lower bounds on $\Gamma_{\rm s}$ as well as $\theta_{\rm s}$.
Thus for example, the radio counterpart of
GRB0508 had a flux density of $0.61\pm0.04\mJy$
at $8.46\GHz$ $6.2$~days after the burst (\cite{IAU6662}).
Using a smooth interpolation between the ultrarelativistic and
nonrelativistic formulae above, and for the relation between
$\Gamma_{\rm s}$ and $\Gamma_{\rm ps}$, we find
$\beta_{\rm s}\gtrsim 0.5$ and $\theta_s\gtrsim 1.5\micas$ 
at the time of this observation, assuming $z=1$ and
$d=10^{28}\cm$ ($3\Gpc$).
The spherical blastwave models cited above predict somewhat larger values,
depending on parameters.
For example, \cite{Sari}'s estimates imply
$\Gamma_{\rm s}\approx 2.6 (E_{52}/n_1)^{1/8}$ at $t=6.2\dy$,
where $E_{52}$ is the energy of the blastwave in units of
$10^{52}~\mbox{erg}$ and $n_1$ is the density of the external
medium in $\cm^{-3}$; this implies $\theta_{\rm s}\approx
9\micas$ for the same values of $d$ and $z$ as above.
``Jet'' models, in which the ejecta are funneled
into a small solid angle, predict angular sizes $\sim 4$ times
smaller than corresponding spherical ones but 
may already be ruled out because
they vastly underpredict the observed radio flux (\cite{Rhoads}).

Thus, if GRB0508 proves to be typical,
it is unlikely that radio afterglows will be directly resolved by
VLBI.
An upper limit of $300\micas$ has been set by the VLBA
eight days after the burst (\cite{TBF}).

Afterglows may, however, be resolved indirectly
by interstellar scattering.
Inhomogeneities in the density of free electrons ($N_e$) affect
the refractive index ($n$) governing the propagation of radio
waves through the interstellar medium:
$\delta n = -(r_e\lambda^2/2\pi)\delta N_e$, where $r_e\equiv e^2/m_e c^2$,
and $\lambda$ is the wavelength.
Extensive evidence supports the conclusion that these inhomogeneities
span a broad range of spatial scales (at least $10^9$ to $10^{14}\cm$)
with a power-law power spectrum 
\begin{equation}
\Phi_{N_e}(\bq)\equiv (2\pi)^{-3}
\int d^3\bs \langle\delta N_e(\bx)\delta N_e(\bx+\bs)\rangle
e^{i\bq\cdot\bs} = C_N^2 q^{-11/3},
\label{eq:spectrum}
\end{equation}
where $\bq$ is the spatial wavenumber conjugate to separation
$\bs$ and is measured in radians per unit length (cf. \cite{ARS}).
Small-scale atmospheric density fluctuations
have power spectra of the same mathematical form, so that radio
scintillation has much in common with optical seeing
(\cite{Coulman}).

Scintillation modulates the flux
in two conceptually different ways (cf. \cite{Rickett90}):

(i) {\it Refractive} scintillation, the random focusing
and defocusing of rays,
can be understood and analyzed entirely within geometric optics.
Given a turbulent spectrum such as (\ref{eq:spectrum}),
the strength of refractive scintillation varies smoothly
with the angular size of the source (inhomogeneities
smaller than the projected source size have no effect because
surface brightness is conserved) and with wavelength
($\delta n\propto\lambda^2$).
Refractive scintillation is insensitive to the observing
bandwidth.

(ii) {\it Diffractive} scintillation is a physical-optics
effect; it can be explained as the
interference among multiple paths from source to receiver.
Since it requires the interfering rays to be mutually coherent,
diffractive scintillation is quenched when the angular
size of the source or the bandwidth of observation is too large.
Diffractive scintillation leads to a
Rayleigh distribution of flux, with fluctuations equal to the
mean: $P(F)=\exp(-F/\langle F\rangle),$
$\langle F^2\rangle=2\langle F\rangle^2$.

Among common radio sources, only pulsars are compact enough to
display diffractive scintillation,
but refractive scintillation affects interstellar masers
and extragalactic sources as well.

In \S 2, we consider diffractive scintillation of radio afterglows.
We find that this is just
possible but very sensitive to the actual angular size of
the afterglow and to the radio frequency of observation.
Detection of diffractive scintillation would strongly constrain
the Lorentz factor of the burst.
In \S 3, we describe the more robust
predictions for refractive scintillation, which can also be
used to constrain the angular size.
Flux variations relative to the mean 
$\approx 0.1(\theta_{\rm s}/\micas)^{-7/6}\nu_{10}^{-2}$ 
are predicted on timescales $\sim 10\hr$.
A summary and some suggestions for
observing strategies are presented in \S 4.

\section{DIFFRACTIVE SCINTILLATION}

For convenience, we imagine in this section only
that the scattering medium is
compressed onto a plane perpendicular to the line
of sight at distance $z_{\rm sc}$ (``thin-screen approximation'').
The characteristic deflection due to scattering is
\begin{eqnarray}
\theta_{\rm d} &=& 2.341\lambda^{11/5} r_e^{6/5}(SM)^{3/5}
\nonumber\\
&=& 2.93 \nu_{10}^{-11/5}
\left(\frac{SM}{10^{-3.5}\m^{-20/3}\kpc}\right)^{3/5}\micas
\label{eq:thetad}
\end{eqnarray}
Here $SM\equiv\int C_N^2( z)d z$, where $ z$ is a coordinate
along the line of sight, is called the
{\it scattering measure}.  We have scaled it by
a value typical for extragalactic sources at high galactic 
latitudes (\cite{Spangler}).

The precise definition of $\theta_{\rm d}$ involves
the {\it phase structure function}, $D_\phi(s)$,
which is the mean-square phase difference accumulated
along parallel lines of sight separated by transverse
distance $s$.
It follows from eq.~(\ref{eq:spectrum}) that 
$D_\phi(s)=(s/s_{\rm d})^{5/3}$ for an appropriately chosen
constant $s_{\rm d}$.   Then $\theta_{\rm d}\equiv (ks_{\rm d})^{-1}$, 
where $k\equiv 2\pi/\lambda$ is the wavenumber.
Apart from a numerical factor, $s_{\rm d}$ is equivalent to the
Fried parameter $r_0$ of optical seeing.

The scatter-broadened image acts somewhat as a Michelson
stellar interferometer, with a characteristic spacing between
the arms $a\sim z_{\rm sc}\theta_{\rm d}$.
Fringes develop on the observer's plane
if the angular size of the source is smaller than the
resolution of the interferometer, $\theta_a= (ka)^{-1}$.
Diffractive scintillation is caused by the motion of the
observer through this fringe pattern.
Therefore, diffractive scintillation is seen only if
\begin{equation}
\theta_{\rm s} < (k z_{\rm sc}\theta_{\rm d})^{-1}
= 2.25\nu_{10}^{6/5}z_{\rm sc,\kpc}^{-1}(SM_{-3.5})^{-3/5}\micas,
\label{eq:slim}
\end{equation}
where $SM_{-3.5}\equiv SM/(10^{-3.5}\m^{-20/3}\kpc)$.
Condition (\ref{eq:slim}) can be satisfied if the
source size is close to the lower limit discussed in \S 1.

In order to see diffractive flux variations of relative
order unity, however, it is also necessary that
$D_\phi(r_{\rm F})> 1$, or equivalently that $s_{\rm d}< r_{\rm F}$,
where $r_{\rm F}\equiv \sqrt{d_{\rm scr}/k}$ is the Fresnel length
(cf. \cite{Rickett90}).
The rays of geometric optics have an intrinsic width
$\sim r_{\rm F}$, and in order to create interference effects,
the phase fluctuations must be strong enough to create
distinct rays.
If this condition is not fulfilled, there is only
{\it weak scintillation} (weak modulation of the flux):
a good optical site is in the weak-scintillation regime,
so that the stars overhead do not twinkle.
The condition for strong scintillation in the present case is
\begin{equation}
\nu < 10.4 (SM_{-3.5})^{6/17} z_{\rm scr,\kpc}^{5/17} \GHz.
\label{eq:nustrong}
\end{equation}
At the maximum frequency allowed by 
eq.~(\ref{eq:nustrong}), the source-size limit becomes
\begin{equation}
\theta_{\rm s}(\nu_{\max})< 
2.35(SM_{-3.5})^{-3/17}z_{\rm scr,\kpc}^{-11/17} \micas.
\label{eq:opt}
\end{equation}
Intriguingly, this is comparable to the minimum source size 
estimated in \S 1 on physical grounds.
Both the scattering measure $SM$ and the effective distance 
$z_{\rm scr}$ scale as $\csc b$ with galactic latitude
[$b({\rm GRB0508})\approx 27^\circ$].
Hence the {\it maximum} measurable size (\ref{eq:opt}) is
$\propto (\sin b)^{14/17}$.

Future afterglows may therefore be indirectly resolvable
by diffractive scintillation when they occur at moderately
high latitude, especially if they can be detected shortly
after the burst ($\lesssim 1\dy$).
Scintillation will be quenched at some frequency
below (\ref{eq:nustrong}) because of the $\nu^{6/5}$
dependence of the constraint (\ref{eq:slim}), and the
transition frequency will provide an indirect measure
of the source size.
The scattering measure along the line of sight can
be internally estimated from the scintillation timescale,
\begin{equation}
t_{\rm diff}= \frac{s_{\rm d}}{v_\perp}=
3.1 \nu_{10}^{6/5} (SM_{-3.5})^{-3/5}
\left(\frac{v_\perp}{30\kms}\right)^{-1}\hr.
\label{eq:tdiff}
\end{equation}

In very strong scintillation, diffractive flux variations
are uncorrelated at radio frequencies differing by more than
the {\it decorrelation bandwidth}
\begin{equation}
\Delta\nu_{\rm dc} = \frac{c}{2\pi\theta_{\rm d}^2d_{\rm scr}}
\approx 7.6\nu_{10}^{22/5}(SM_{-3.5})^{-6/5}d_{\rm scr,\kpc}^{-1}\GHz.
\label{eq:nudc}
\end{equation}
As this formula shows, the frequencies of interest are not far from
the borderline between strong and weak scintillation, where
$\Delta\nu_{\rm dc}\sim\nu$.
On the one hand, this is fortunate because the afterglows
are faint and must be observed with a broad bandwidth.
On the other hand, it is unfortunate because $\Delta\nu_{\rm dc}$
measures an independent combination of $SM$ and $d_{\rm screen}$.

The angular size limit (\ref{eq:slim}) and the decorrelation
bandwidth (\ref{eq:nudc}) become more severe as the distance
to the scattering screen increases,
and it might therefore be supposed that scattering in the
gamma-ray burst's host galaxy (if it exists) or in the
intergalactic medium (IGM) would suppress diffractive scintillation.
At worst, however, such scattering has the effect of replacing
one incoherent source (of size $\theta_{\rm s}$) with an effective
incoherent source of somewhat larger angular size $\hat\theta_{\rm s}$.
Diffractive scintillation due to the Galactic interstellar medium (ISM)
will still occur provided that $\hat\theta_{\rm s}$ satisfies the limit 
(\ref{eq:slim}) with $d_{\rm scr}$ determined by the path length
through the Galactic ISM.
Also, a deflection angle $\hat\theta_{\rm d}$ in the host galaxy
contributes to the effective source size in proportion to
\begin{displaymath}
\frac{d_{\rm ISM}}{d_{\rm host}}\hat\theta_{\rm d}\sim 10^{-6}
\hat\theta_{\rm d}.
\end{displaymath}
Hence, scattering in the host can probably be neglected, for the
same reason that orbiting satellites looking down enjoy much
better seeing than ground-based telescopes looking up.

Scattering by the IGM is negligible presuming (in the
absence of better information) that the scattering measure
scales as the path length times the square of the mean electron
density: 
\begin{displaymath}
\frac{SM_{\rm IGM}}{SM_{\rm ISM}}\sim\frac{1\Gpc}{1\kpc}
\left(\frac{2\times 10^{-7}\cm^{-3}}{0.02\cm^{-3}}\right)^2\sim 10^{-4}.
\end{displaymath}
We have assumed a fully-ionized IGM containing most of the baryons
allowed by primordial nucleosynthesis 
($\Omega_{\rm b}h_{50}^2=0.05\pm0.01$: \cite{ITL})
and taken the local electron density from \cite{TC}.
In the case of GRB0508,
scattering by an intervening galaxy associated with the 
$z=0.835$ absorption system might be comparable to that due
to the local ISM, but there appears to be no intrinsically bright
galaxy on the line of sight [\cite{IAU6674}], so it is likely
that the absorption comes from the outer halo of a galaxy,
where the electron density and scattering measure are likely to be small.

\section{REFRACTIVE SCINTILLATION}

Following \cite{CFRC} (henceforth CFRC), the normalized spatial
correlation of the flux due to refractive scintillation is
\begin{eqnarray}
C(\bs)&\equiv&
\frac{\langle F_\nu(\bx)F_\nu(\bx+\bs)\rangle}{\langle F_\nu\rangle^2}~-1
= 8\pi r_e^2\lambda^2\int\limits_0^\infty
dz\int d^2\bq\, e^{i\bq\cdot\bs}\Phi_{N_e}(q_x,q_y;q_z=0;z)\nonumber\\
&\times&\left|V\left(\frac{\bq  z}{k}\right)\right|^2
\exp\left[-\int\limits_0^\infty d\bar z 
D'_\phi\left(\frac{\bq z_<}{k},\bar z\right)\right]
\sin^2\left(\frac{q^2  z}{2k}\right).
\label{eq:Cint}
\end{eqnarray}
Here the line of sight runs from the observer at $ z=0$
to the source at (effectively) $ z=\infty$.
The electron-density spectrum $\Phi_{N_e}$ [eq.~(\ref{eq:spectrum})]
depends on $z$ via the coefficient $C_N^2\to C_N^2(z)$, which
reflects changes in the strength of the turbulence along the line
of sight.
We take $C_N^2( z)= C_N^2(0)\exp[-( z\sin b/H)^2]$,
where $C_N^2(0)$ is the local value near the Sun, and 
$ z\sin b$ is the height above the plane.
The flux correlation depends upon the average (or sum) of the
phase fluctuations along the line of sight, hence the $z$ component
of $\bq$ is set to zero in eq.~(\ref{eq:Cint}).

The quantity $V({\br})$ is the
visibility of the source on baseline ${\br}$, which
like $\bq$ and $\bs$ in this formula, is a vector
transverse to the line of sight.
We assume a gaussian image brightness distribution for the source,
$I_\nu(\vec\theta)\propto\exp(-\vec\theta^2/2\theta_s^2)$,
so that $V(\bq  z/k)=\exp(-q^2 z^2\theta_s^2/2)$.

Finally,
\begin{eqnarray}
D'_\phi(\br,z)&\equiv& 4\pi r_e^2\lambda^2\int d^2\bq
\Phi_{N_e}(q_x,q_y;q_z=0,z)\left[1-e^{i\bq\cdot\bs}\right],\nonumber\\
&=&2^{4/3}\frac{3\pi^2\Gamma(7/6)}{5\,\Gamma(11/6)}C_N^2(z)
r_e^2\lambda^2 s^{5/3}
\label{eq:struct}
\end{eqnarray}
is the differential phase structure function.
In other words, $D'_\phi(\br,z)dz$  is the contribution of the
slab $(z,z+dz)$ to the mean-square phase difference
between lines of sight separated by $\br$.
The quantity $z_<\equiv\min(z,\bar z)$.

The three final factors in equation (\ref{eq:Cint}) can
be regarded as the squares of ``visibilities'' due to the
intrinsic source size, to the scattering, and to
Fresnel optics.
The characteristic angular scales associated with 
these visibilities at $z=H\csc b$ are
$\theta_{\rm s}$, $\theta_d$, and  the {\it Fresnel angle}
\begin{equation}
\bar\theta_{\rm F}\equiv (kH\csc b)^{-1/2}
= 2.57 \nu_{10}^{-1/2}\left(\frac{H\csc b}{\kpc}\right)^{-1/2}\micas.
\label{eq:thetaF}
\end{equation}
Actually, the Fresnel visibility at $z=H\csc b$ is 
\begin{equation}
V_{\rm F}(\br)\equiv\frac{\sin(k\theta_{\rm F} r/2)}{k\theta_{\rm F} r/2},
\end{equation}
so that $V_{\rm F}(0)=1$.
All three visibilities have the effect of restricting the range
of wavenumbers $\bq$ that contribute to $C(\bs)$.
They occur multiplicatively in equation (\ref{eq:Cint}) because the
effective source size seen by the observer is a convolution of the
intrinsic surface brightness distribution $I_\nu(\vec\theta)$
with the scatter-broadened image of a point source, and with
a sort of Fresnel image.

The root-mean-square variation in the flux relative to its mean value
is $\sqrt{C(0)}\equiv m_R,$ the {\it modulation index}.
From equation (\ref{eq:Cint}),
\begin{eqnarray}
m_R&=&\left[\frac{1}{4}\Gamma(7/6)\Gamma(1/3)\right]^{1/2}
r_e\lambda^2\theta_{\rm eff}^{-7/6}C_N^2(0)^{1/2}
(H\csc b)^{1/3}\nonumber\\
&=&0.12\left(\frac{\theta_{\rm eff}}{10\micas}\right)^{-7/6}
\nu_{10}^{-2}
\left(\frac{H\csc b}{1\kpc}\right)^{1/3}
\left(\frac{C_N^2(0)}{10^{-3.5}\m^{-20/3}}\right)^{1/2},
\label{eq:modex}
\end{eqnarray}
where the effective source size is
\begin{equation}
\theta_{\rm eff}\equiv\left[\theta_{\rm s}^2+(0.7064\theta_{\rm d})^2
+(0.8482\theta_{F})^2\right]^{1/2}.
\label{eq:thetaeff}
\end{equation}
This extends the results of CFRC, who gave explicit formulae 
for $m_R$ in the limit of large intrinsic size,
$\theta_{\rm s}\gg\theta_{\rm d},\theta_{\rm F}$.
Our result (\ref{eq:modex}) is an interpolation formula.
The numerical coefficients in the definition of $\theta_{\rm eff}$
have been chosen so that equation (\ref{eq:modex}) is exact when
any one of the three characteristic angles
$(\theta_{\rm s},\theta_{\rm d},\theta_{\rm F})$ is much larger
than the other two.

Refractive scintillation occurs on 
the timescale required for the line of sight to
cross the image at a mean distance $d_{\rm scr}\approx H\csc b/2$,
\begin{equation}
t_{\rm ref}= \frac{\theta_{\rm eff}H\csc b}{2v_\perp}
=7.0 \left(\frac{\theta_{\rm eff}}{10\micas}\right)
\left(\frac{H\csc b}{\kpc}\right)
\left(\frac{v_\perp}{30\kms}\right)^{-1} ~\mbox{hr},
\label{eq:tref}
\end{equation}
assuming that the motion of the line of sight is
more rapid than the internal velocities of the medium.
The full temporal correlation function is then $C(v_\perp\tau)$
at lag $\tau$.
Note that unlike $t_{\rm diff}$ [eq.~(\ref{eq:tdiff})],
$t_{\rm ref}$ remains constant or decreases with increasing frequency,
depending on the intrinsic image size.  At frequencies above
limit (\ref{eq:nustrong}), where the scintillation is weak,
the two timescales merge into a single one:
the time required for the line of sight to cross the Fresnel length.

In principle, it is possible to estimate the angular size from
the amplitude and timescale of refractive scintillation
at a single frequency.
In practice, it may be difficult to disentangle refractive
flux variations from instrumental and atmospheric noise;
futhermore, the parameters $C_N^2(0)$ and $H$ are imperfectly
known.
Therefore it will be safer to measure $m_R$ at several frequencies.
Equations (\ref{eq:modex}), (\ref{eq:thetaeff}), and (\ref{eq:thetad})
imply that $m_R\propto \nu^{-2}$ at relatively high frequencies, where 
$\theta_{\rm eff}\approx\theta_{\rm s}$ is frequency-independent; and
that $m_R\propto\nu^{+17/30}$ at low frequencies, where
$\theta_{\rm eff}\approx\theta_{\rm d}\propto\nu^{-11/5}$.
Hence $m_R(\nu)$ has a peak:
\begin{eqnarray}
\nu_{\rm peak} &=& 4.30
\left(\frac{\theta_{\rm s}}{10\micas}\right)^{-5/11}
(SM_{-3.5})^{3/11}\GHz,\nonumber\\
m_{R,\rm~peak} &=& 0.270
\left(\frac{\theta_{\rm s}}{10\micas}\right)^{-17/66}
\left(\frac{H\csc b}{\kpc}\right)^{-7/33}
\left(\frac{C_N^2(0)}{10^{-3.5}\m^{-20/3}}\right)^{1/22}.
\end{eqnarray}
These formulae assume that $\theta_{\rm s}$ is 
substantially larger than the Fresnel angle (\ref{eq:thetaF})
at the peak.

\section{Discussion}

If radio afterglows of gamma-ray bursts are indeed
produced by relativistic blastwaves at cosmological distances,
then their small angular size and high
surface brightness puts them in an extremely interesting
part of parameter space with respect to interstellar scintillation.
Both diffractive and refractive scintillation are possible for these
sources.

{\it Diffractive scintillation} is sensitive to the radio frequency
of observation, $\nu$, and the angular size of the source,
$\theta_{\rm s}$.
It may have been marginally possible to have seen diffractive
scintillation in GRB0508 at $\nu\approx 10\GHz$ if 
$\theta_{\rm s}\lesssim 2\micas$ [eq.~(\ref{eq:slim})].
Future afterglows from sources at high galactic latitude
will present other opportunities for observing diffractive
scintillation, which will be recognized because
\begin{itemize}
\item the flux variations are of unit strength, 
$\langle F^2\rangle=2\langle F\rangle^2$ [but contrary to
the situation with pulsars, scintillation will be correlated
accross a broad frequency band, cf. eq.~(\ref{eq:nudc})];

\item the timescale decreases with decreasing radio frequency,
and is on the order of a few hours [eq.~(\ref{eq:tdiff})];

\item the effect vanishes abruptly below a frequency determined
by the scattering measure and the angular size of the source
[eq.~(\ref{eq:nustrong})].
\end{itemize}

Diffractive scintillation will allow the interstellar
medium to be used as a Michelson stellar interferometer to resolve
angular sizes measured in micro-arcseconds, far too small for
conventional VLBI.

{\it Refractive scintillation} will occur at all frequencies and
for all angular sizes, though its amplitude depeneds algebraically
on both of thes parameters.
The timescale will range upward from a few hours.
The amplitude, though less than unity, will be large ($\sim 10\%$)
unless $\theta_{\rm s}$, and hence the Lorentz
factor of the blastwave, are much larger than predicted by current 
models [eq.~(\ref{eq:modex})].
Refractive scintillation therefore can also be used to constrain
$\theta_{\rm s}$ and $\Gamma_{\rm s}$, 
though it will not provide quite so
sharp a test as diffractive scintillation.

Some implications for observing strategies are the following: 

One wants to collect data at several frequencies, roughly in the
range $2-20\GHz$.

For diffractive scintillation, the flux should be sampled on timescales
$\sim 1~\mbox{hr}$, and the first data points should be taken as early
as the sensitivity of the equipment permits, since the source will be
smaller and less likely to be overresolved by the scattering
[see eqs.~(\ref{eq:slim}) \& (\ref{eq:opt})].
One should probably attempt to see diffractive scintillation
in sources at high galactic latitude ($\gtrsim 30^\circ$) only, once
again because these are least likely to be overresolved.
Ideally one would wait for a source close to a lines
of sight on which the scattering parameters have been well determined
from observations of extragalactic sources and globular-cluster
pulsars.

For refractive scintillations,  many of the same considerations
apply.
In this case, however, one can afford to make observations later
in the lifecycle of the afterglow, at longer time intervals, 
and at lower galactic latitude.
It will be important, however, to minimize or accurately subtract
the contribution of receiver and atmospheric noise to the flux
variations, since the amplitude of the modulation index is critical
to the estimate of the angular size.

I am indebted to Michael Rupen, Andrew Ulmer, Michael
Richmond, Bohdan Paczy\'nski, and the rest of the Observatory coffee
klatsch for stimulating discussions and for criticism of the 
first draft.
Special thanks are owed to Michael Richmond for help with 
the IAU circulars and for his ephemeris software.

\end{document}